\documentclass[preprint,proceedings]{rmaa}

\usepackage{paralist}

\usepackage{psfrag,color}

\SetYear{2008}
\SetConfTitle{Astronom\'{\i}a Din\'{a}mica en Latinoam\'{e}rica}

\title{The Anomalous Acceleration of the Pioneer Spacecrafts}

\author{Jos\'{e} A. de Diego\altaffilmark{1}}

\altaffiltext{1}{Instituto de Astronom\'{\i}a, Universidad Nacional
  Aut\'{o}noma de M\'{e}xico, Ciudad Universitaria,
  Apdo Postal 70-264, 04510 M\'{e}xico D.F., M\'{e}xico
  (jdo@astroscu.unam.mx).}

\shortauthor{De Diego}
\shorttitle{Acceleration of the Pioneer}

\listofauthors{J. A. de Diego}
\indexauthor{De Diego, J. A.}

\abstract{Radiometric data from the Pioneer 10 and 11 spacecrafts have revealed an unexplained constant acceleration of $a_A = (8.74 \pm 1.33) \times 10^{-10} \rm{m\,s^{-2}}$ towards the Sun, also known as the Pioneer anomaly. Different groups have analyzed the Pioneer data and have got the same results, which rules out computer programming and handling errors. Attempts to explain this phenomenon arguing intrinsic causes on-board the spacecrafts failed or have lead to inconclusive results. Therefore, the Pioneer anomalous acceleration has motivated the interest of researchers to find out explanations that could bring insight upon the forces acting in the outer Solar Systems or a hint to discover new natural laws.}

\resumen{Los datos radiom\'{e}tricos de las naves Pioneer 10 y 11 han revelado una aceleraci\'{o}n constante no explicada de $a_A = (8.74 \pm 1.33) \times 10^{-10} \rm{m\,s^{-2}}$ dirigida hacia el Sol, tambi\'{e}n conocida como la anomal\'{\i}a del Pioneer. Distintos grupos han analizado los datos de los Pioneer y han obtenido los mismos resultados, lo que descarta errores en los c\'{o}digos de computadora y en el manejo de los datos. Los intentos por explicar este fen\'{o}meno argumentando causas intr\'{\i}nsecas a bordo de las naves fracasaron o han conducido a resultados no concluyentes. Debido a esto, la aceleraci\'{o}n an\'{o}mala del Pioneer ha motivado el inter\'{e}s de los investigadores por encontrar explicaciones que puedan echar luz sobre las fuerzas que act\'{u}an en el Sistema Solar exterior o una pista para descubrir nuevas leyes naturales.}

\addkeyword{Gravitation}
\addkeyword{Space vehicles}
\begin{document}
% Typeset article header
\maketitle

\section{Introduction}
\label{sec:intro}

During the last decade, two unexplained phenomena have been discovered that alter the predicted course of spacecrafts and satellites. One is known as the Pioneer anomaly, a constant acceleration towards the Sun detected in spacecrafts traveling in the outer Solar System, and which is the subject of this review. The other phenomenon is named the Flyby anomaly \citep{and01}, which consists in a sudden, unexpected velocity increase by a few $\rm{mm \; s^{-1}}$ first observed during the Earth flyby of the Galileo spacecraft on December 8, 1990. This anomaly has shown again during the flybys of the Near Earth Asteroid Rendezvous (NEAR) spacecraft on January 23, 1998 \citep{ant98}, and during the flybys of Cassini-Huygens (August 18, 1999), Rosetta (March 4, 2005) and Messenger \citep[August 2, 2005;][]{and07}. A recent analysis of the flybys of all these spacecrafts can be found in \citet{and08}. Although both Pioneer and Flyby anomalies might have a prosaic explanation, they might also be the clue to unreveal fundamental laws of the nature.

The navigation system of the Pioneer 10 and 11 spacecrafts is the most precise aboard of any deep space vehicle up to date. This navigation system was designed to support high precision experiments in celestial mechanics. Hence, the Pioneers have a mHz precision Doppler tracking with an acceleration sensitivity of $10^{-10}\, \rm{m\,s^{-2}}$, an advanced spin-stabilized attitude control, and Radioisotope Thermoelectric Generator (RTGs) attached on extended arms that contribute to the spacecraft stability and to reduce heat systematics \citep{and98}. These capabilities allowed the detection of an anomalous constant acceleration $a_A = (8.74 \pm 1.33) \times 10^{-10} \rm{m\,s^{-2}}$ towards the Sun \citep{and98}, that has also been suggested in the radiometric data from Galileo, Ulysses and Cassini spacecrafts \citep{and02,and03}. The acceleration cannot be imputed to failures in the tracking algorithm, and both engineering causes and external forces have been invoked but none possible explanation has been confirmed yet.

The trajectories and ultimately the fate of the five spacecrafts are very different. Thus, Pioneer 10 and 11 follow approximate opposite escape hyperbolic trajectories close to the plane of the ecliptic, and they will at long last reach the Oort Cloud and abandon the Solar System. For Galileo, the spacecraft crashed into Jupiter on September 21, 2003. Ulysses has flown over the Sun's poles for the third time in 2007 and 2008; as its aging radioisotope generators continue to run down the mission is coming to an end after 18 years. Cassini remains orbiting around Saturn and its mission will be extended probably until July 2010. The presence of the same anomalous effect despite the differences in trajectories and physical design among these spacecrafts may be a hint of the existence of external forces or unknown physical laws. However, on-board causes cannot be ruled out and they are intensively studied to find an explanation.

\begin{figure}[!t]
    \parbox[t]{\textwidth}{
     \vspace{20pt}
  \includegraphics[width=0.5\columnwidth]{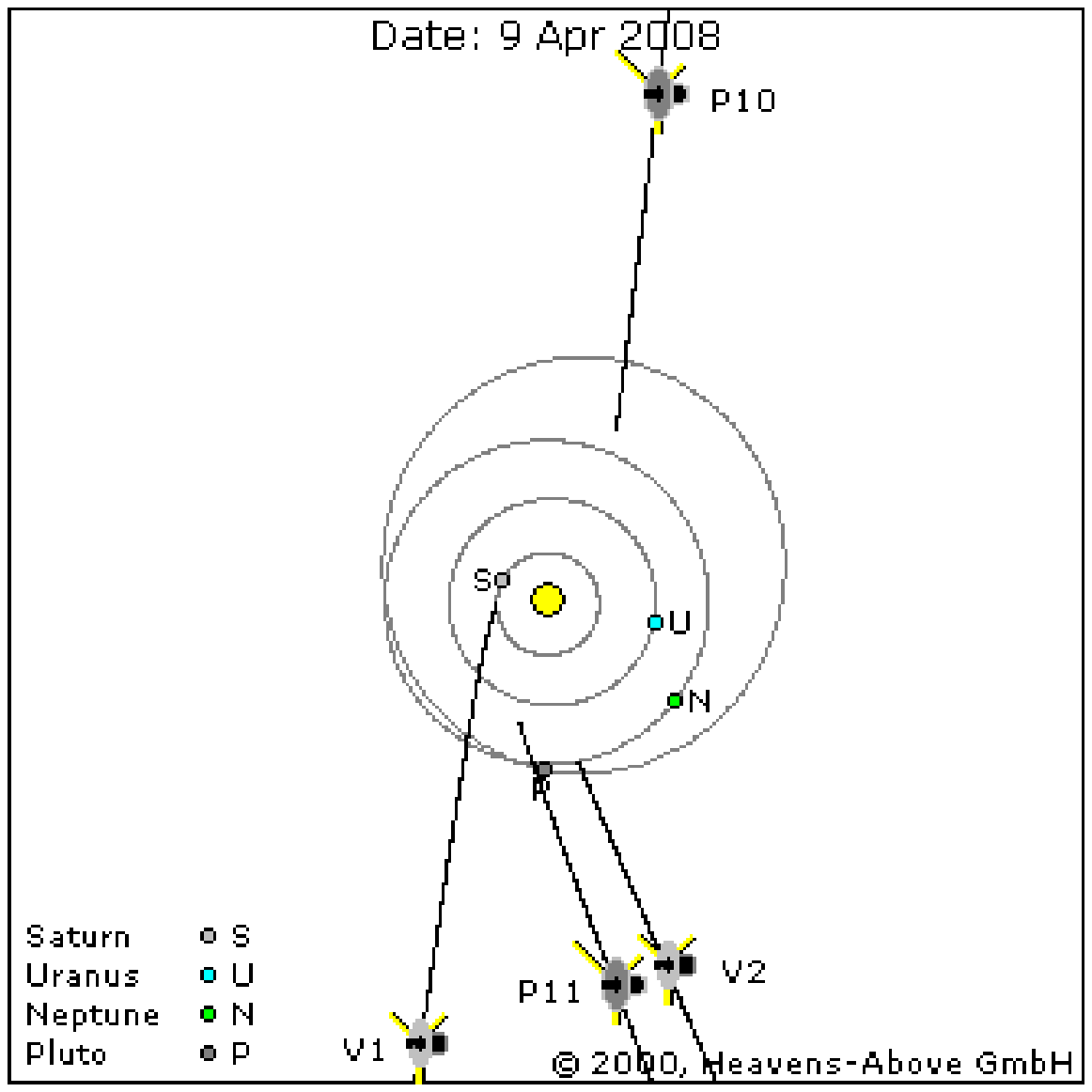}
  \includegraphics[width=0.5\columnwidth]{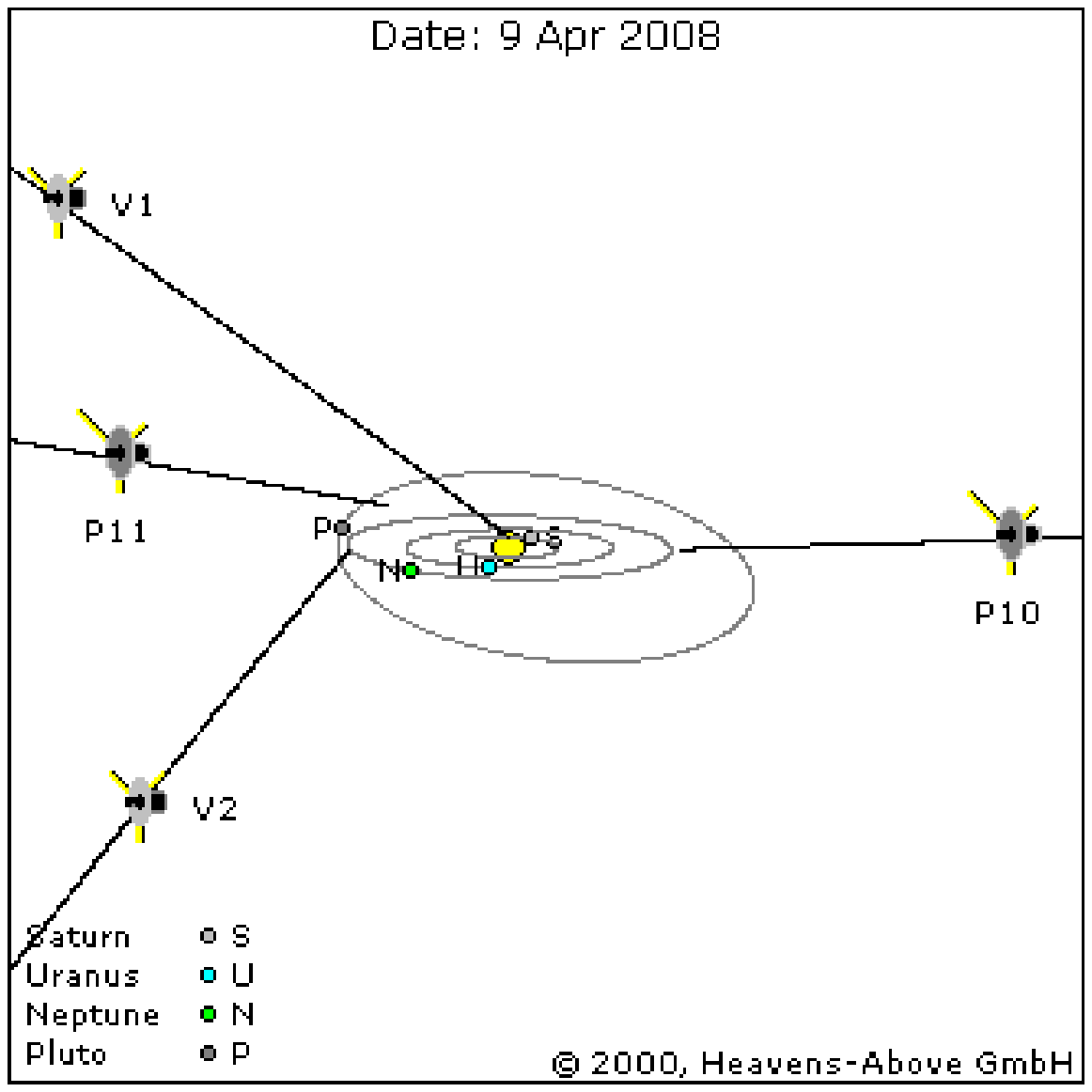}
  }
  \caption{Pioneers 10 and 11, and Voyagers 1 and 2 positions and trajectories up to April 9, 2008 (Reproduced from Heavens Above: http:$/ \! /$www.heavens-above.com).}
  \label{fig:trajectories}
\end{figure}

Pioneer 10 and 11 were launched on March 2, 1972 and April 5, 1973, respectively. Pioneer 10 was the first spacecraft to travel through the asteroid belt and encountered Jupiter (1973) and Pluto(1983), while Pioneer 11 visited also Jupiter (1974) and explored the planet Saturn (1979). Pioneer 10 is currently within the Sun's heliopause. The power source of the Pioneer 10 degraded during the mission affecting the strength of its signal, while the Pioneer 11 antenna is misaligned and the spacecraft cannot be operated to point back at the Earth.
The last communications with Pioneer 10 and 11 occurred in January 2003 and November 1995, respectively. As for April 9, 2008, Pioneer 10 was about 95.47 AU from the Sun, and Pioneer 11 around 75.90 AU; their positions, as well as for Voyager 1 and 2, are shown in Fig.~\ref{fig:trajectories}.

This work is organized as follows. Section~\ref{sec:acceleration} poses the Pioneer anomaly. In Section~\ref{sec:intrinsic} we review the on-board causes that have been invoked to explain the anomaly. Section~\ref{sec:extrinsic}  presents a summary of the external forces that have been proposed to act on the spacecrafts. In Section~\ref{sec:future} we discuss possible next steps to accomplish in the research of what causes the anomaly. Finally, Section~\ref{sec:conclusions} summarizes the conclusions of this paper.

\section{The unmodeled acceleration}
\label{sec:acceleration}

\begin{figure}[!t]
  \includegraphics[width=\columnwidth]{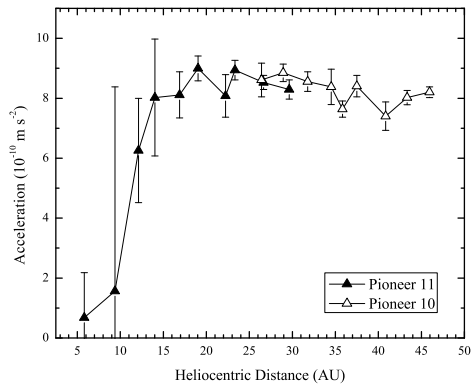}
  \caption{Unmodeled accelerations on Pioneer 10 and 11. The acceleration starts near Uranus, around 20 AU, but the onset of the perturbation may have started near Saturn, around 10 AU. Figure adapted from \citet{and02}.}
  \label{fig:acceleration}
\end{figure}

The Pioneer trajectories have been modeled from radiometric data, considering gravitational and non-gravitational forces acting upon the spacecrafts. Starting around 20 AU from the Sun, near the orbit of Uranus, the models deviate from the radiometric data by a small Doppler frequency blue-shifted drift of $(5.99 \pm 0.01) \times 10^{-9} \, \rm{Hz \, s^{-1}}$ \citep{and98,and02}. \citet{and02} interpreted this Doppler drift either as a constant, unexplained acceleration towards the Sun of $(8.74 \pm 1.33) \times 10^{-10} \rm{m\,s^{-2}}$, or a constant time deceleration of $(2.92 \pm 0.44) \times 10^{-18} \, \rm{s\,s^{-2}}$. Fig.~\ref{fig:acceleration} shows the unmodeled acceleration; note that the onset of the anomaly may start around 10 AU, namely around the orbit of Saturn.

The origin of the anomaly has been attributed by different authors either to spacecraft intrinsic or extrinsic causes, which will be reviewed in the next sections. Other possibilities are a non calibrated bias in the data, a spurious result due to the approximation algorithms to calculate the orbits and  statistics, or an error introduced by the navigational software used to calculate the Pioneer trajectories. However, four independent studies with different softwares have confirmed that the presence of the anomaly is not an artifact introduced in the calculations \citep{tur06}\footnote{Turyshev raises the number of independent studies up to seven (Turyshev, S.~G.\ 2007, The Planetary Society, http:$/ \! /$ www.planetary.org/programs/projects/innovative\_technologies \\ /pioneer\_anomaly/update\_20070328.html).}.

\section{Intrinsic causes}
\label{sec:intrinsic}

\begin{figure}[!t]
  \includegraphics[width=\columnwidth]{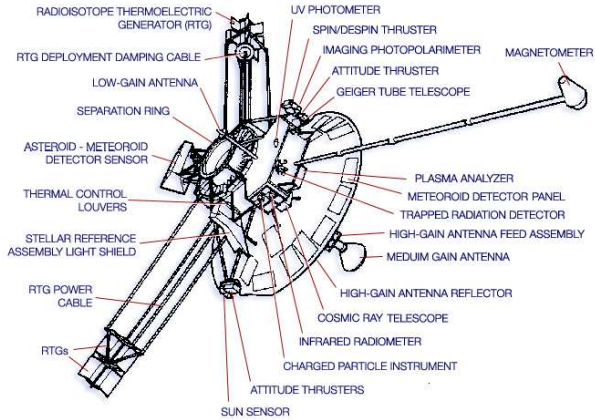}
  \caption{Pioneer 10 and 11 spacecrafts (Reproduced from NASA, http:$/ \! /$ www2.jpl.nasa.gov/basics/pioneer.html).}
  \label{fig:pioneer}
\end{figure}

The external parts of the Pioneer spacecraft are identified in Fig.~\ref{fig:pioneer}. A summary of the intrinsic effects proposed to explain the Pioneer Anomaly is presented in  Table~\ref{tab:intrinsic}.

\citet{and02,and02b} consider several spacecraft intrinsic causes that may be responsible for the Pioneer anomaly. These effects include electromagnetic emission from the spacecrafts (either thermal losses or beamed from the spacecraft antennae) and  gas leaks.

\begin{table*}[!t]\centering
  \tablecols{4}
  % Stretch the space between table columns
  \setlength{\tabcolsep}{2\tabcolsep}
  \caption{intrinsic effects} \label{tab:intrinsic}
  \begin{tabular}{llr}
    \toprule
    Responsible & \multicolumn{1}{c}{Effects} & \multicolumn{1}{c}{Acceleration} \\
    & & \multicolumn{1}{c}{($10^{-10} \; \rm{m \; s^{-2}}$)} \\
    \midrule
    Antenna & Radio signal & $1.10\pm0.10$ \\
    RTGs & Antenna reflection & $-0.55\pm0.55$ \\
    RTGs & Anisotropicemission & $\pm0.85$ \\
    RTGs & He expulsion & $0.15\pm0.16$ \\
    Propulsion system & Gas leakage & $\pm0.56$ \\
    Electrical circuits & Electrical heat & Non constant \\
    \bottomrule
  \end{tabular}
\end{table*}

The largest effect may originate in the recoil force of the radio beam emitted from the antenna. The acceleration due to the emission of the radio signal may account for an acceleration of $(1.1 \pm 0.1) \times 10^{-10} \; \rm{m\;s^{-2}}$. However, as the antenna aims to the Earth, the force exerted by the radio beam is in the opposite direction to the discovered acceleration, which would become even larger.

\citet{and98,and02} point out that the heat emitted from the RTGs, which is anisotropically reflected by the spacecraft high gain antennae, may also contribute to the observed anomaly. These authors estimate that the heat reflection may account for $(-0.55{\pm}0.55) \times 10^{-10} \; \rm{m \; s^{-2}}$ of the observed anomaly. An independent estimate by \citet{sch03} rises the heat emitted by the RTGs and hence the acceleration adds up to $-3.3 \times 10^{-10} \; \rm{m \; s^{-2}}$. Note that the radiant heat from the RTGs decreases by the decay of the radioactive fuel, and thus the amount of the acceleration should decrease with time. However, \citet{ols07} investigates the temporal variations of the anomalous acceleration and concludes that the Pioneer 10 and 11 Doppler data is not accurate enough to distinguish between a constant acceleration and acceleration proportional to the remaining plutonium in the RTGs.

Dependency of the surface degeneration of the RTGs on to the spacecrafts orientation has also been considered by \citet{and02} as a possible cause of the anomaly. The inner sides of the RTGs received the solar wind during the early parts of the missions, while the outer sides received the impact of the Solar System dust particles. Both effects can degrade the surfaces of the RTGs and produce asymmetries in the heat radiated away from the RTGs in the fore and aft directions. \citet{and02} estimated an upper limit for the contribution to the anomalous acceleration due to the uncertainty of the asymmetric emissivity, which amounts to $0.85 \times 10^{-10} \; \rm{m \; s^{-2}}$. However, this mechanism also depends on the radioactive decay, and thus a decrease of the acceleration would be expected, in discordance with the observed constancy of the anomaly.

\citet{mur99} investigated the heat produced by the electrical power, which may be redirected through the closed thermal control louvers \citep{sch03}, but the resulting acceleration would also decrease with the radioactive decay.

As a result of the $\alpha$-decay of $^{238}$Pu, the RTGs hold some quantity of He. The anomaly could be explained if this He gas escapes in one direction at a tiny rate of $0.77 \; \rm{g \; yr^{-1}}K$, but \citet{and02} have ruled out this mechanism and estimated its contribution to the anomaly as $(0.15 \pm 0.16) \times 10^{-10} \; \rm{m \; s^{-2}}$.

\citet{and02} have also estimated the acceleration uncertainty imputable to gas leakage from the propulsion system ($0.56 \times 10^{-10} \; \rm{m \; s^{-2}}$), but it is unlikely that such a random mechanism would affect both Pioneers producing the same outcome.

\section{Extrinsic causes}
\label{sec:extrinsic}

The difficulty to find an intrinsic effect that can produce the same observed acceleration in both Pioneer 10 and 11, and that the same anomaly is suggested for Galileo, Ulysses and Cassini spacecrafts, despite their different designs and trajectories, make the research to find an explanation based on external forces acting on the spacecrafts very attractive. In subsection \ref{sec:conventional} we review some of the conventional external forces that have been proposed as a mechanism of acceleration on the spacecrafts. More hypothetical explanations, which rely on less established theories, are commented in \S~\ref{sec:newphysics}.

\subsection{Conventional forces}
\label{sec:conventional}

Several nongravitational, conventional forces have been proposed by different authors to explain the Pioneer anomaly. Hence, drag force due to interplanetary dust have been investigated by \citet{nie05} and \citet{ber06} who calculated that the density of dust necessary to provoke the acceleration would be five orders of magnitude larger than the density calculated for the Kuiper belt dust ($\sim10^{24} \; \rm{g \; cm^{-3}}$). \citet{bin04} discuss a nongravitational acceleration of the Sun, orthogonal to the ecliptic, but they found that it is necessary that the Sun would emit all the electromagnetic radiation in the opposite direction. \citet{lam06} have studied the coincidence of the Pioneer anomalous acceleration with the value cH, where c is the velocity of light and H the Hubble constant, and the possible influences on the signal propagation, trajectory of the spacecraft, magnitude of the gravitational field and the definition of the astronomical unit due to the cosmic expansion; however, these authors calculate that the effect can only account for a value of vH, where v is the spacecraft, i.e. a factor v/c less than the observed anomalous acceleration velocity. An origin related to the cosmological expansion has also been proposed by \citet{oli07}. This author conjectures that the Solar System has escaped the gravity of the Galaxy as evidenced by its orbital speed and radial distance and by the visible mass within the solar system radius. Spacecrafts unbound to the solar system would also be unbound to the galaxy and subject to the Hubble law.

A gravitational source in the Solar System as a possible origin for the anomaly has been considered by \citet{and02}. According to the equivalence principle, such a gravitational source would also affect the orbits of the planets. In the case of the inner planets, which have orbits determined with great accuracy, they show no evidence for the expected anomalous motion if the source of the anomaly were located in the inner Solar System. For example, in the case of Mars, range data provided by the Mars Global Surveyor and Mars Odyssey missions have yielded measurements of the Mars system center-of-mass relative to the Earth to an accuracy of one meter \citep{kon06}. However, the anomaly has been detected beyond 20 AU (i.e., beyond Uranus, 19 AU), and the orbits of the outer planets have been determined only by optical methods, resulting in much less accurate planet ephemerides.

Attempts to detect observable evidence of unexpected gravitational effects acting on the orbits of the outer planets have not yield any positive results yet. Hence, \citet{rat06} used parametric constraints to the orbits of Uranus and Neptune and found that the reduced Solar mass to account for the Pioneer anomaly would not be compatible with the measurements. A similar result was obtained by \citet{ior06} based on the Gauss equations to estimate the effect of a gravitational perturbation in terms of the time rate of change on the osculating orbital elements. These authors argue that the perturbation would produce long-period, secular rates on the perihelion and the mean anomaly, and short-period effects on the semimajor axis, the eccentricity, the perihelion and the mean anomaly large enough to be detected. \citet{tan07} also considers the effect on the path of the outer planets by a disturbance on a spherically symmetric space-time metric, and rules out any model of the anomaly that implies that the Pioneer spacecrafts move geodesically in a perturbed space-time metric. A recent test for the orbits of 24 Trans-Neptunian Objects using bootstrap analysis also failed to find evidence of the anomaly in these objects \citep{wal07}.

Nevertheless, the evidence for non perturbation of the orbits in the outer Solar System is weak and unconclusive, as there are not range data measurements yet. For example, \citet{pag06} conclude that such anomalous gravitational disturbance would not be detected in the orbits of the outer planets, and recently \citet{sta08} discuss modifications to the laws of gravitation that can explain the anomaly and still be acceptable to the ephemerides of the planets from Saturn and outward. Therefore, efforts to find a gravitational explanation continue as in the case of a recent paper by \citet{nya08} who proposes an azimuthally symmetric solution to Poison's equation for empty space to explain qualitatively the Pioneer anomaly. This solution results in a gravitational potential dependent on the distance and the polar angle, and it has also implications for the planetary orbits albeit they are not tested with ephemeris data yet.

\begin{figure}[!t]
  \includegraphics[width=\columnwidth]{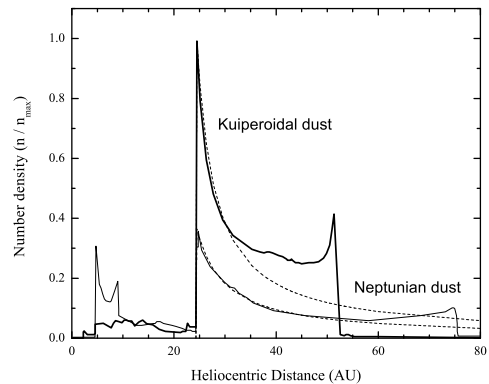}
  \caption{Numerical simulation of the radial density distribution of the Solar System matter at the ecliptic plane, normalized for the number density distribution of the kuiperoidal dust. Solid broad line shows the kuiperoidal dust; solid thin line shows the neptunian dust; dashed lines exemplifies $\rho \propto (R-k)^{-1}$ laws for these belts analogous to the profile distribution for the Solar System dark matter proposed by \citet{die06}. Profiles for the kuiperoidal and neptunian dust adapted from \citet{gor98}.}
  \label{fig:kuiperdust}
\end{figure}

The possibility of a gravitational perturbation on the Pioneer paths has been also considered by \citet{nie05} and \citet{ber06}, who studied the possible effects produced by different Kuiper Belt mass distributions, and they conclude that the Kuiper Belt cannot produce the observed acceleration. \Citet{die06} also discarded the gravitational attraction by the Kuiper Belt, but they suggest that the observed deceleration in the Pioneer space probes can be simply explained by the gravitational pull of a distribution of undetected dark matter in the Solar System. Thus, considering a NFW dark matter distribution \citep{nav97}, \citet{die06} show that there should be several hundreds earth masses of dark matter available in the Solar System. \citet{gor98} has shown that the Solar System dust distributes in two dust systems and four resonant belts associated with the orbits of the giant planets. As shown in Fig~\ref{fig:kuiperdust}, the density profile of these belts approximately follows an inverse heliocentric distance dependence law ($\rho \propto (R-k)^{-1}$, where $k$ is a constant). As in the case of dark matter, dust is usually modeled as a collisionless fluid because the pressure force is negligible. Although dust is subjected to radiation pressure, this effect is very small in the outer Solar System. Given this similarity, for \citet{die06} a spatial distribution of part of the Solar System dark matter analogous to these dust belts could explain the observed anomaly. The dark matter gravitational pull has been recently considered also by \citet{nie08} who proposes the analysis of the New Horizons spacecraft data when the probe crosses the orbit of Saturn (see \S~\ref{sec:future}).

Pioneer anomaly explanations involving dark matter depend on the small scale structure of NFW haloes, which is not known. Hence, N-body simulations to investigate Solar System size subhalos would require of the order of $10^{12}$ particles \citep{nat05}, while the largest current simulations involve around $10^8$ particles \citep[e.g. the Via Lactea model consists of 234 million particles][]{die07}. As a consequence
of this lack of knowledge on the dark matter small scale structures, the existence of a dark matter halo around the Sun is still an open question. Thus, it has been proposed that the dark matter could become trapped in the Sun's gravitational potential after experiencing multiple scatterings \citep[e.g.][]{pre85}, perharps combined with perturbations due to the planets \citep{dam98}. Moreover, the Solar System itself may be a consequence of the existence of a local halo. The existence of dark matter streams crossing the Solar System, perhaps forming ring-shaped caustics analogous to the dark matter ring postulated by \citet{die06}, has been also considered by \citet{sik98}.

Lately, \citet{and07} have investigated the spacecraft acceleration in terms of the flyby anomaly, and they find a possible onset of the Pioneer effect as a result of the Jupiter (for Pioneer 10) and Saturn (for Pioneer 11) flybys.

\subsection{New physics}
\label{sec:newphysics}

Many explanations involving modifications of the gravitation theory or hypothetical effects ascribable to a number of physical agents have been proposed. In this paragraph we summarize some of these efforts which reflect the richness of the debate forced by the lack of a definitive conventional explanation for the Pioneer anomaly.

The possibility of an unknown interaction of the Pioneer radio signals with the solar wind was considered by \citet{and98}. A hypothetical class of dark matter that would restore the parity symmetry, the so-called \emph{mirror matter}, has been considered by \citet{foo01}. Modified-inertia approaches have been considered under the Modified Newtonian Dynamics theory  \citep{mil01} and by Unruh radiation \citep{mcc07}. \citet{cad04} studied the coupling of gravity with a scalar field with an exponential potential, while \citet{ber04} also applied a scalar field in braneworld scenarios. \citet{jae05} presents a solution in terms of the parameterized post-Newtonian formalism. Gravitational coupling resulting in an increase of the constant G with scale is analyzed by \citet{ber96}. \citet{mof04} discusses the anomaly in terms of a nonsymmetric gravitational theory. \citet{ran04} investigates the effect of a background gravitational potential that pervades the universe and is increasing because of the expansion, provoking a drift of clocks \citep[see also][]{and02}; however, such an effect should also be observed in the radio signals from pulsars \citep{mat97,wex01}, which is not the case. \citet{ost02} proposes that cosmic expansion applies directly to gravitationally bound systems according to the so-called quasi-metric framework. According to \citet{ros04,ros05}, the scale factor of the space-time background would cause an anomaly in the frequency. The cosmological constant has also been invoked to produce acceleration by \citet{not08} and a gravitational frequency shift by \citet{mbe04}. Finally, a number of possible tests of general relativity in the Solar System have been recently reviewed by \citet{rey08}.

\section{Future research}
\label{sec:future}

There have been a number of interesting proposals to launch a mission to investigate the Pioneer anomaly. Hence, \citet{dit05} propose a dedicated mission based on a formation-flying approach that consists of an actively controlled spacecraft and a set of passive test-masses. On the other hand, \citet{rat06} propose a non-dedicated mission consisting either of a planetary exploration spacecraft or a piggybacked micro-spacecraft to be launched from a mother spacecraft traveling to Saturn or Jupiter. Several challenging technological goals have been visualized for such missions, such as positioning control, thermal design, control of the antennae emission, etc. A precision 2-3 orders of magnitude better than the Pioneer spacecrafts (i.e., around $10^{-12} \; \rm{m \; s^-2}$) would be necessary.

A dedicated mission to study the anomaly cannot help being very risky until the possibility that on-board effects have been completely ruled out. Nevertheless, the possibilities that would open the discovery of an external influence or the breakthrough of a new physical law are so fascinating that any future Solar System mission should have the capabilities to test the anomaly. In the meantime, it would be worth to use current Solar System missions as anomaly probes. In this respect, a recent proposal by \citet{nie08} consists in the analysis of the data from the New Horizons spacecraft traveling to Pluto and the Kuiper-Belt. The spacecraft was launched on January 19, 2006, and on its pass through the orbit of Saturn in mid-2008 could supply a clear test of the onset of a Pioneer-like anomaly, as suggested by the Pioneer data. In the future, an increase in the accuracy on the position and velocities of the comets (possibly landing probes with telemetric capabilities on their surfaces) would permit to test the external effects on their motion within large regions of the Solar System.

Previous to any dedicated space mission to study the Pioneer anomaly, it is absolutely essential to analyze the complete Pioneer database in order to rule out, as much as possible, any on-board cause. In this sense, a remarkable effort to rescue and analyze early Pioneer data (before 1987) is currently in progress \citep{tot08}. These data include telemetric measurements as well as the physical state of the spacecraft instruments (temperatures, currents and voltages, gas pressure). A careful analysis of the thermal and gas losses from the spacecrafts will be very useful in modeling and testing intrinsic possible causes for the anomaly, while the early telemetric data possibly will bring new light on the onset of the anomaly. For example, these data might be very important to discriminate the direction of the perturbation, and thus the possible origin of the anomaly. Therefore, if the anomaly is directed to the Sun it would suggest a Solar or Solar System origin; if directed to the Earth it would be probably associated with the frequency standards; if the anomaly is in the direction of the spacecraft motion, it would be related to inertial or drag forces; and if the anomaly is linked with the direction of the rotational axis, it would be a strong evidence for intrinsic spacecraft causes.

\section{Conclusions}
\label{sec:conclusions}

The Pioneer Anomaly consists in an unmodeled constant acceleration of $a_A = (8.74\pm1.33) \times 10^{-10} \; \rm{m \; s^{-2}}$ towards the Sun detected in radiometric data from the Pioneer 10 and 11 spacecrafts, and also suggested in the radiometric data from Galileo, Ulysses and Cassini spacecrafts. Although there have been many efforts to disentangle its nature, the anomaly still has an uncertain origin. Spurious results introduced by the approximation algorithms, as well as errors in the navigational software used to calculate the trajectories of the Pioneers, have been ruled out after four independent studies have proven evidence of the same anomalous effect. Heat radiation or gas leaks and other on-board causes cannot be completely ruled out, but it is tough to uphold that the same intrinsic effect shows up in five spacecrafts that differ both in designs and trajectories. This circumstance has stimulated the search for extrinsic causes that can explain the anomaly. Hence, different researchers have argued about gravitational disturbances and other conventional forces acting upon the spacecrafts. Although dark matter in the outer Solar System may be a strong candidate to explain the anomaly, it is extremely difficult to prove its effect on the orbits of the planets beyond Saturn.

Another line of research has been held by various groups in the sense that the observed anomaly might be a result of the incompleteness of the current theory of gravitation, or even an indication of new physical phenomena. As speculative as this line of research is, it is undoubtedly very attractive because it can uncover new clues and unexpected physical laws. Scientists are eager to confront unexplained phenomena and therefore new attempts will arise to formulate a solid explanation for the Pioneer anomaly. Perhaps, the mystery of the Pioneer anomaly will not be resolved until a space mission especially devoted to investigate the dynamics in the outer Solar System collects accurate enough data.

\acknowledgements

This work has been supported by the CONACyT grant 50296. Updated positions for the Pioneer 10 and 11, and Voyager 1 and 2, have been obtained from Heavens Above (http://www.heavens-above.com). The picture describing the Pioneer spacecraft has been obtained from NASA (http:$/ \! /$ www2.jpl.nasa.gov/basics/pioneer.html). This work has made use of NASA's Astrophysics Data System Bibliographic Services.

\end{document}